\documentclass[prd,twocolumn,aps,showpacs,preprintnumbers,amsmath,amssymb]{revtex4}
\usepackage{graphicx,graphics,color}% Include figure files
\usepackage{dcolumn}% Align table columns on decimal point
\usepackage{bm}% bold math
\usepackage{psfrag}% psfrag

\begin{document}
%\draft
\newcommand{\be}{\begin{equation}}\newcommand{\ee}{\end{equation}}
\newcommand{\bea}{\begin{eqnarray}}\newcommand{\eea}{\end{eqnarray}}
\newcommand{\bc}{\begin{center}}\newcommand{\ec}{\end{center}}
\def\no{\nonumber}
\def\eq#1{Eq. (\ref{#1})}\def\eqeq#1#2{Eqs. (\ref{#1}) and  (\ref{#2})}
%%%%%%%%%%%%%%%%%%%%%%%%%%%%%%%%%%%%%%%%%%%%%%%%%%%
\def\lsim{\raise0.3ex\hbox{$\;<$\kern-0.75em\raise-1.1ex\hbox{$\sim\;$}}}
\def\gsim{\raise0.3ex\hbox{$\;>$\kern-0.75em\raise-1.1ex\hbox{$\sim\;$}}}
\def\slash#1{\ooalign{\hfil/\hfil\crcr$#1$}}
\def\eff{\mbox{\tiny{eff}}}
\def\order#1{{\mathcal{O}}(#1)}
\def\etp{\eta^{\prime}}\def\etetp{\eta^{(\prime)}}
\def\bketp{B\to K \etp}\def\bket{B\to K \eta}\def\bketetp{B\to K \eta^{(\prime)}}
\def\bkzetp{B^0\to K^0 \etp}\def\bkzet{B^0\to K^0 \eta}\def\bkzetetp{B^0\to K^0 \eta^{(\prime)}}
\def\bk#1#2{\langle 0|#1|#2 \rangle}\def\la{\langle}\def\ra{\rangle}
%%%%%%%%%%%%%%%%%%%%%%%%%%%%%%%%%%%%%%%%%%%%%%%%%%%
\preprint{CP3-06-15}
\title{Anomalous enhancement of a  penguin hadronic matrix element in $\bketp$}
\author{
J.-M. Gerard\footnote{email address: gerard@fyma.ucl.ac.be} and 
E. Kou\footnote{email address: ekou@fyma.ucl.ac.be}} 
\address{Institut de Physique Theorique and 
 Centre for Particle Physics and Phenomenology (CP3)\\
Universit\'{e} Catholique de Louvain, 
Chemin du Cyclotron 2, 
B-1348 Louvain-la-Neuve, Belgium
} 
\date{\today}
\begin{abstract}
We estimate the density matrix element for the $\pi^0, \eta$ and $\etp$ production from the vacuum in the large-$N_c$ limit.  
As a consequence, we find  that the QCD axial anomaly leads to highly
non-trivial corrections to the usual flavour SU(3) relations
between 
$B^0 \to K^0\pi^0$, $B^0 \to K^0 \eta$ and  $B^0 \to K^0 \eta^{\prime}$ decay amplitudes. 
These corrections may  explain why
the $B\to K \etp$ branching ratio is about six times larger than the
$B\to K \pi$ one.
\end{abstract}
\maketitle
%%%%%%%%%%%%%%%%%%%%%%%%%%%%%%%%%%%%%%%%%%%%%%%%%%
First observations of a large branching ratio for $\bketp$ triggered numerous theoretical investigations within and beyond the standard model. The latest average of the available experimental data~\cite{hfag}: 
\begin{equation}
Br(\bkzetp)= (64.9\pm 3.5)\times 10^{-6}
\label{eq:1}
\end{equation}
definitely confirms a sizable excess of $\etp$ compared with $\pi^0$, $Br(B^0\to K^0\pi^0)=(10.0\pm 0.6)\times 10^{-6}$, and $\eta$, $Br(B^0\to K^0\eta)<1.9\times 10^{-6}$. The fact that  $\etp$ is  mostly a flavour $SU(3)$-singlet naturally suggests mechanisms such as the so-called singlet contribution ($\etp$ production by  gluon-gluon fusion from $b\to sg$ or $b\to sgg$)~\cite{singlet} or intrinsic charm  contribution ($\etp$ production by $c\bar{c}$ annihilation from $b\to c\bar{c}s$)~\cite{charm}. However, these  do not seem to explain why the $\bkzetp$ branching ratio is about six times larger than the $B^0\to K^0\pi^0$ one.  More recently, the hadronic parameters for $\etp$ production have been  re-examined and a couple of different solutions were proposed~\cite{BN, zupan,Li}. In this letter, we analyse an overlooked  correction  from the axial  $U(1)$ anomaly in the  hadronic matrix elements associated with $b\to s\bar{d}d, s\bar{s}s$. 

The derivative of the flavour-singlet axial current is given by: 
\be
\partial_{\mu}j^{\mu 5}_0 =2i\sum_{q=u,d,s}m_q\bar{q}\gamma_5q+\frac{3\alpha_s}{4\pi} G_{\alpha\beta}\tilde{G}^{\alpha\beta}\label{eq:deriv}
\ee
where $G_{\alpha \beta}$ is the  gluonic field strength tensor and $\tilde{G}^{\alpha\beta}$,  its dual. 
The mass term in the r.h.s. of Eq. (\ref{eq:deriv}) implies explicit flavour $SU(3)$ violation. This breaking alone would lead to the {\it ideal} mass relations 
\be (M_{\eta}^2)_{\tiny ideal}=2M_K^2-M_{\pi}^2,  
\quad (M_{\etp}^2)_{\tiny ideal}=M_{\pi}^2  \label{eq:ideal}
\ee
 which are quite successful for the $(\phi, \omega, \rho)$ vector mesons but totally unrealistic for the $(\eta, \etp, \pi)$ pseudoscalar mesons~\cite{Weinberg}. 
However, the second term in the r.h.s. of Eq. (\ref{eq:deriv}) represents the QCD anomaly  which breaks the axial $U(1)$ symmetry  to provide the $\etp$ with a mass around 1 GeV~\cite{tHooft}. 
 
 The hadronic matrix element $\bk{G_{\alpha\beta}\tilde{G}^{\alpha\beta}}{\etetp}$ associated with the anomalous term in Eq. (\ref{eq:deriv}) has been considered in relation to  the Zweig-suppressed $(J/\psi \to \etetp \gamma)$ radiative decays. In this letter, we aim at an estimate of the corresponding hadronic matrix elements for the first term in the r.h.s. of Eq. ({\ref{eq:deriv}). This will then allow us to compute the contribution of the dominant penguin density-density operator to the $(B^0\to  K^0 \pi^0, \eta, \etp)$ hadronic decays in a way fully consistent with the axial $U(1)$ and flavour $SU(3)$ symmetry-breaking requirements.  

In a naive quark picture, the hadronic matrix elements $\bk{\bar{q}\gamma_5q}{\etp, \eta, \pi^0}$ are simply related through Clebsch-Gordan coefficients (C.G.). These coefficients are fixed by the $q\bar{q}$ content in the pseudoscalar wave-functions. However, corrections due to flavour $SU(3)$ but also to  axial $U(1)$ violations have to be taken into account. In particular, we expect the following 
generic form: 
\be
\left.\frac{\bk{\bar{q}\gamma_5q}{\etetp}}{\bk{\bar{q}\gamma_5q}{\pi^0}}\right|_{q=u,d} = \mbox{ C.G.} \big\{1+\frac{M_{\etetp}^2-M_{\pi}^2}{\Lambda^2}\big\}. \label{eq:guess}
\ee
The appearance of the physical $\etp$ mass may, at first sight, be surprising. But this is in fact required by the axial $U(1)$ symmetry. 
Indeed, in the absence of the axial anomaly, the {\it ideal} mass relations 
given in Eq. (\ref{eq:ideal}) would consistently imply that the $\etp/\pi^0$ ratio in Eq (\ref{eq:guess}) is equal to $+1$ if $q=u$ and $-1$ if $q=d$ since the  $\etp$ and $\pi^0$ wave-functions have the same quark content as the iso-singlet $\omega$ and the iso-triplet
$\rho$ in this fictitious  world. The axial anomaly calls thus for a sizable correction to the $\bk{\bar{q}\gamma_5q}{\etp}$ hadronic matrix elements  if the cut-off scale is what we naturally expect from the real QCD dynamics, namely $\Lambda=\order{1}$ GeV. So, these matrix elements have to be consistently extracted from a low-energy effective theory of QCD. 
  
Again with reference to the observed pattern for the pseudoscalar mass spectrum, let us consider the large-$N_c$ limit at each order in the (squared) momentum $p^2$. The genuine $U(3)_L\times U(3)_R$ chiral invariant structure of the effective non-linear theory implies then the following hierarchy~\cite{GK}:   
\be
\order{p^0,1/N_c} > \order{p^2,1/\infty} > \order{p^4,1/\infty} \cdots \label{eq:hierarchy}
\ee  
such that the full pseudoscalar mass spectrum naturally arises in three steps. The leading ($p^0$) term ensures the breaking of the $U(1)_A$ symmetry and provides the flavour-singlet $\eta_0$ with a mass around 1 GeV. The next-to-leading ($p^2$) term implies the usual $SU(3)_V$ mass splitting among the ($\pi, K, \eta_8$) flavour-octet. Eventually, next-to-next-leading ($p^4$) terms are needed to precisely reproduce the observed $\eta-\etp$ mass splitting.   

Starting from this effective theory of QCD in the large-$N_c$ limit, we may consistently express the anomalous operator of Eq. (\ref{eq:deriv}) purely in terms of the flavour-singlet field $\eta_0$. This allows us to extract the $\eta-\etp$ mixing angle associated with the diagonalization of the $\eta_8-\eta_0$ squared mass matrix: 
\bea
\eta& =& \eta_8 \cos\theta -  \eta_0\sin\theta \nonumber \\
\etp&=& \eta_8\sin\theta+\eta_0\cos\theta \label{eq:2-2-3}
\eea
from the well-measured ($J/\psi \to \etetp\gamma$) radiative decays~\cite{GK}. The extracted value is 
$\theta=-(22\pm 1)^{\circ}$. Let us emphasise once again that in the absence of the  axial anomaly, $\theta$ would have been equal to the {\it ideal} octet-singlet mixing angle $+35.3^{\circ}$ of the $\phi-\omega$, i.e.
\be
\cos\theta_{\tiny ideal} =\sqrt{2/3}, \quad 
\sin\theta_{\tiny ideal}=\sqrt{1/3}. \label{eq:idealm}
\ee
since the ideal mass relations given in Eq. (\ref{eq:ideal}) correspond to the
wave-functions $\eta=-s\bar{s}$ and $\etp=(u\bar{u}+d\bar{d})/\sqrt{2}$. 
This illustrates how $\etetp$ masses
and mixing are strongly correlated through the $U(1)_A$ symmetry.

Similarly, we may also express the density operator of Eq. (\ref{eq:deriv}) in terms of the flavour-nonet field $\pi$: 
\bea
&&\bar{q}^a\gamma_5 q^{b} \supset i \frac{f\ r}{2\sqrt{2}} \\
&& \ \ \ \ \ \ \times\left[
1-\frac{\partial_{\mu}\partial^{\mu}}{\Lambda_0^2}+r(m_a+m_b)\left(\frac{1}{\Lambda_1^2}-\frac{1}{4\Lambda_2^2}\right)\right] \pi^{ba}\no
\eea
where   $a, b=1,2,3$ are the flavour indices.  
The parameters $r$ and $f$ are related to the physical masses and decay constants as  
\bea
&M_{\pi}^2=r m_q[1+M_{\pi}^2(\frac{2}{\Lambda_1^2}-\frac{1}{\Lambda_2^2})] &\\
&M_K^2=\frac{r (m_q+m_s)}{2}[1+M_K^2(\frac{2}{\Lambda_1^2}-\frac{1}{\Lambda_2^2})]&
\eea
and 
\bea
&f_{\pi}=f[1+M_{\pi}^2(\frac{1}{\Lambda_0^2}+\frac{1}{2\Lambda_2^2})] &\\
&f_{K}=f[1+M_{K}^2(\frac{1}{\Lambda_0^2}+\frac{1}{2\Lambda_2^2})] &
\eea
if isospin symmetry is assumed, $q= u \mbox{ or } d$. 
As a result, we obtain the following set of density matrix elements:  
\bea
&\bullet&\la0|  \bar{q} \gamma_5q | \pi^0  \ra =\pm i\frac{f_\pi}{2\sqrt{2}}\frac{M_{\pi}^2}{m_q}\label{eq:r-pi} \\
&\bullet&\la0|  \bar{q}\gamma_5 s | K  \ra =i{f_K}\frac{M_{K}^2}{m_s+m_q} \label{eq:r-k} \\
&\bullet&\la0|  \bar{q}\gamma_5 q | \eta  \ra = \label{eq:r-1}\\
& &i\frac{f_\pi}{2\sqrt{6}}\frac{M_{\pi}^2}{m_q}\left(c\theta -\sqrt{2}s\theta\right) \left[1+\frac{M_{\eta}^2-M_\pi^2} {\Lambda_0^2}\right]\nonumber \\
&\bullet&\la0|  \bar{q}\gamma_5 q | \etp  \ra = \label{eq:r-2}\\
&&i\frac{f_\pi}{2\sqrt{6}}\frac{M_{\pi}^2}{m_q}\left(\sqrt{2}c\theta +s\theta\right)\left[1+\frac{M_{\eta^{\prime}}^2-M_\pi^2} {\Lambda_0^2}\right] \nonumber \\ 
&\bullet&\la0|  \bar{s}\gamma_5 s | \eta  \ra = - i\frac{f_K}{\sqrt{3}}\frac{M_{K}^2}{m_s+m_q}\left(\sqrt{2}c\theta +s\theta\right)\label{eq:r-3} \\
&&\ \ \ \times \left[1+\frac{M_{\eta}^2-M_K^2}{\Lambda_0^2}+2(M_K^2-M_\pi^2)\left(\frac{1}{\Lambda_1^2}-\frac{1}{4\Lambda_2^2}\right)\right] \no \\ 
&\bullet&\la0|  \bar{s}\gamma_5 s | \etp  \ra =  i\frac{f_K}{\sqrt{3}}\frac{M_{K}^2}{m_s+m_q}\left(c\theta -\sqrt{2}s\theta\right)\label{eq:r-4} \\
&&\ \ \ \times \left[1+\frac{M_{\etp}^2-M_K^2}{\Lambda_0^2}+2(M_K^2-M_\pi^2)\left(\frac{1}{\Lambda_1^2}-\frac{1}{4\Lambda_2^2}\right)\right] \no  
\eea
where we introduce the abbreviation $(c\theta, s\theta)$ for $(\cos\theta, \sin\theta)$. On the basis of Eq. (\ref{eq:hierarchy}), we neglect $\order{1/\Lambda_i^4}$ corrections. 
This parametrisation of the density matrix elements for $\pi^0, \eta$ and $\etp$ follows from a $U(3)_L\times U(3)_R$ invariant theory and is thus fully consistent with the $U(1)_A$ and $SU(3)_V$ symmetry requirements on the pseudoscalar mixing {\it and} masses  in the isospin limit. 

Eqs. (\ref{eq:r-pi}) and (\ref{eq:r-k}) are the well-known hadronic matrix elements already derived in the large-$N_c$ limit using chiral perturbation theory~\cite{chiE}. These hadronic matrix elements feel the usual effects of $SU(3)_V$ violation on the decay constants and masses in the pseudoscalar flavour-octet. 

 Eqs. (\ref{eq:r-1}) and (\ref{eq:r-2}) display the highly non-trivial effect of the $U(1)_A$ breaking: in the absence of the axial anomaly, the {\it ideal} masses (see Eq. (\ref{eq:ideal})) and mixing (see Eq. (\ref{eq:idealm})) would consistently  imply $\bk{\bar{q}\gamma_5q}{\eta}=0$ and $\bk{\bar{q}\gamma_5q}{\etp}=\pm\bk{\bar{q}\gamma_5q}{\pi^0}$. They nicely confirm our  original guess expressed in Eq. (\ref{eq:guess}). 
 Moreover, a global fit of  
the pseudoscalar masses, mixing and  decays constants  has already been undertaken  in our previous work~\cite{GK}. The resulting values for the $\Lambda_{0,1,2}$ cut-offs: 
\be
\Lambda_0 \simeq 1.2 \ \mbox{GeV},  \ \ \ \Lambda_1\simeq 1.2\ \mbox{GeV}, \ \ \ \Lambda_2\simeq 1.3\ \mbox{GeV} \label{eq:v6-13}
\ee
are indeed all around one GeV, as anticipated. 

Eqs. (\ref{eq:r-3}) and (\ref{eq:r-4}) consistently combine the effects of $U(1)_A$ and $SU(3)_V$ violations. Our  results for the scale-independent density matrix elements $2m_s \la 0|\bar{s}\gamma_5s|\etetp \ra$ are compared with previous works  in Table 1. 
\begin{table*}[t!]
\begin{center}
\scalebox{1.2}{\begin{tabular}{|c|c|c|c|c|}
\hline 
&\multicolumn{2}{|c|}{this work} & \multicolumn{2}{|c|}{previous works} \\
\hline 
$\theta=-22^{\circ}$& $U(1)_A\times SU(3)_V$  & $SU(3)_V$ & AG~\cite{AG}& BN~\cite{BN} \\
\hline\hline 
$2im_s \la 0|\bar{s}\gamma_5s|\eta \ra$ & $+0.053 \pm 0.008$ &$+0.058$& $+0.057$& $+0.055$\\ 
\hline
$2im_s \la 0|\bar{s}\gamma_5s|\etp \ra$ &$-0.109 \pm  0.016$&$-0.069$ &$-0.071$&$-0.068$ \\
\hline
%$\zeta$& 1.11 &1.19& 1.25 & 1.20 \\
%\hline
%$\zeta^{\prime}$ &1.53&1.08&1.10&1.06\\
%\hline
\end{tabular}}
\caption{Comparison of numerical results for the $\bar{s}\gamma_5s$  density matrix elements in GeV$^3$ units. The second column includes flavour $SU(3)$ breaking with a realistic octet-singlet  mixing angle, $\theta=-22^{\circ}$, but with the {\it ideal} mass relations for $\etetp$  (see Eq. (\ref{eq:ideal})). Our result agrees with the previous works in this peculiar limit. The first column includes full $U(1)_A$ and $SU(3)_V$ breaking effects. We find that the magnitude of the $\etp$ hadronic matrix element increases by about 60\%  while the one for $\eta$ only decreases by about 10\%. }
\end{center}
\end{table*}
The second column includes flavour $SU(3)$ breaking with a realistic octet-singlet  mixing angle  but with the {\it ideal} mass relations for $\etetp$ (see Eq. (\ref{eq:ideal})). 
We can see an excellent numerical agreement in this limit which is
rather peculiar since any realistic $\eta-\etp$ mixing excludes ideal
$\eta-\etp$ masses from the viewpoint of the $U(1)_A$ symmetry. 
 The first column includes full $U(1)_A$ and $SU(3)_V$ breaking effects. The magnitude of the $\etp$ hadronic matrix element increases then by about 60\% while the one for $\eta$ only decreases  by about 10\%. 
%This large correction for  $\bk{\bar{s}\gamma_5s}{\etp}$ should not come as a surprise if one realizes from Eq. (\ref{eq:idealm}) that this hadronic matrix element exactly vanishes in the absence of the axial
%anomaly! 
Theoretical uncertainties associated with higher order
$SU(3)_V$ and $U(1)_A$ corrections are expected to be  $M_K^4/\Lambda^4\simeq 5$\% and  $M_K^2M_{\etetp}^2/\Lambda^4\simeq 15$\%, respectively.

Eqs. (\ref{eq:r-pi}) to (\ref{eq:r-4}) turn out to be crucial for an estimate of the $\bkzetetp$ decay amplitudes. Indeed, their typical $M^2/m$ chiral enhancement, at the basis of the $\Delta I=1/2$ rule in $K\to \pi\pi$ decays~\cite{BBG}, is such that the  
 $Q_6\equiv \sum_{q}(\bar{b}_Lq_R)(\bar{q}_Rs_L)$ weak operator provides the main contribution for the $b\to s\bar{q}q$ penguin-induced hadronic processes involving two pseudoscalars in the final state.  
 (Notice that such is not the case for the vector-pseudoscalar final state, $B\to K^* \etetp$.) 
 In the limit where, diagrammatically, 
  $\bkzetetp$ come from $b\to s\bar{d}d$ and $b\to s\bar{s}s$  penguins, while $B^0\to K^0\pi^0$ comes only from $b\to s\bar{d}d$, we may thus write  
\bea
&&\left. \frac{A(\bkzetetp)}{A(B^0\to K^0\pi^0)}\right|_{Q_6} \label{eq:q6}\\
&=& \frac{ \la K^0 |\bar{b}_Ls_R|{B^0}\ra}{\la \pi^0 |\bar{b}_Ld_R|{B^0}\ra}
\left[\frac{\la \etetp |\bar{b}_Ld_R|{B^0}\ra}{ \la K^0 |\bar{b}_Ls_R|{B^0}\ra}+\frac{\la\etetp|\bar{s}_Rs_L|0\ra}{\la K^0|\bar{d}_Rs_L|0\ra}\right]. \no
\eea
We already notice that in the absence of the axial anomaly, the $\etp$
wave-function has no $s\bar{s}$ component and the $b\to s\bar{s}s$ penguin
contribution to $B^0\to K^0\etp$ vanishes. In that {\it ideal} limit, the
$\etp/\pi^0$ ratio in Eq.(\ref{eq:q6}) is equal to $-1$ and thus $Br(\bkzetp) =
Br(B^0\to K^0\pi^0)$!

\begin{figure*}[t]
\begin{center}
\psfrag{u1}[c][l][1]{}
%\begin{minipage}{4cm}$U(1)$ breaking correction to hadronic matrix element\end{minipage}}
\psfrag{arr}[c][c][.8]{\begin{picture}(2,4)\put(0,0){\vector(0,1){12}}\end{picture}}
\psfrag{arrp}[c][c][.8]{\begin{picture}(2,4)\put(0,0){\vector(0,1){22}}\end{picture}}
\psfrag{x}[c][c][1]{$\zeta^{\prime}$}
\psfrag{y}[c][c][1]{$Br(\bkzet)/Br(B^0\to K^0 \pi^0)$}
\psfrag{yp}[c][c][1]{$Br(\bkzetp)/Br(B^0\to K^0 \pi^0)$}
\psfrag{exp}[c][c][1]{}
\psfrag{no}[c][c][1.1]{$\bullet$}\psfrag{yes}[c][c][2]{$\star$}
\psfrag{a}[c][c][1.2]{\rotatebox{23}{\bf /}}\psfrag{b}[c][c][1.2]{\rotatebox{23}{\bf /}}
\psfrag{noc}[c][c][0.9]{$SU(3)_V$}\psfrag{yesc}[c][c][0.9]{$U(1)_A\times SU(3)_V$}
\psfrag{1}[c][c][1]{1}\psfrag{1.2}[c][c][1]{1.2}\psfrag{1.3}[c][c][1]{1.3}\psfrag{1.4}[c][c][1]{1.4}\psfrag{1.5}[c][c][1]{1.5}\psfrag{1.6}[c][c][1]{1.6}\psfrag{1.7}[c][c][1]{1.7}\psfrag{1.8}[c][c][1]{1.8}\psfrag{2.2}[c][c][1]{2.2}
\psfrag{1}[c][c][1]{1}\psfrag{2}[c][c][1]{2}\psfrag{3}[c][c][1]{3}\psfrag{4}[c][c][1]{4}\psfrag{5}[c][c][1]{5}\psfrag{6}[c][c][1]{6}\psfrag{7}[c][c][1]{7}\psfrag{8}[c][c][1]{8}\psfrag{10}[c][c][1]{10}
\psfrag{0.2}[c][c][1]{0.2}\psfrag{0.175}[c][c][1]{}\psfrag{0.15}[c][c][1]{0.15}\psfrag{0.125}[c][c][1]{}
\psfrag{0.1}[c][c][1]{0.1}\psfrag{0.075}[c][c][1]{}\psfrag{0.05}[c][c][1]{0.05}\psfrag{0.025}[c][c][1]{}
\includegraphics[width=8cm]{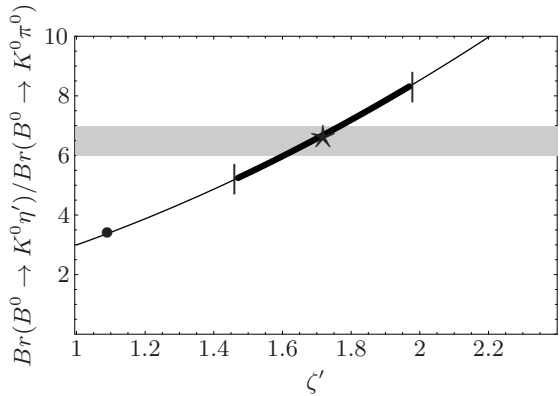}\hspace*{1cm}
\caption{
The $\bkzetp$ to $B^0\to K^0\pi^0$  ratio of branching ratios as a
function of the $\etp$ mass-dependent parameter $\zeta^{\prime}$, for a fixed value
of the mixing angle $\theta = -22^{\circ}$. 
The usual flavour $SU(3)$ relation, $Br(\bkzetp)/Br(B^0\to K^0\pi^0)\simeq 3$, is obtained at $\zeta^{\prime}=1$. The dot corresponds to  $SU(3)_V$ but no $U(1)_A$ corrections to $\zeta^{\prime}$, i.e. $\zeta^{\prime}=1.09$. 
Finally, the star represents full $SU(3)_V$ and $U(1)_A$ corrections to $\zeta^{\prime}$, i.e. $\zeta^{\prime}=1.72\pm 0.26$.  The shaded area
displays the current experimental range.
}
\label{fig:1}
\end{center}
\end{figure*}

As a first order approximation, the effect of the $U(1)_A$ and $SU(3)_V$
violations on the pseudoscalar masses can be safely neglected compared to the $B$-mass scale. So, we simply express the $B$-to-light meson hadronic matrix elements in terms of C.G. and focus on the axial $U(1)$ and flavour $SU(3)$ breaking corrections for the vacuum-to-light meson transitions. In this approximation, we easily obtain:  
\bea
&&\frac{A(\bkzet)}{A(B^0\to K^0\pi^0)}  \label{eq:eta}\\
&=&-\sqrt{2}\left[(\sqrt{\frac{1}{6}}c\theta-\sqrt{\frac{1}{3}}s\theta)+(-\frac{2}{\sqrt{6}}c\theta-\sqrt{\frac{1}{3}}s\theta)\zeta\right]\no \\
&&\frac{A(\bkzetp)}{A(B^0\to K^0\pi^0)} \label{eq:etp}\\
 &=& -\sqrt{2}\left[(\sqrt{\frac{1}{6}}s\theta+\sqrt{\frac{1}{3}}c\theta)+(-\frac{2}{\sqrt{6}}s\theta+\sqrt{\frac{1}{3}}c\theta)\zeta^{\prime}\right]\no
\eea
where the $U(1)_A$ and $SU(3)_V$ breaking effects associated with the pseudoscalar masses are fully encoded in the parameters  
\be
\zeta^{(\prime)}\equiv  1+\frac{M_{\etetp}^2-M_{K}^2}{\Lambda_0^2}+2(M_K^2-M_{\pi}^2)(\frac{1}{\Lambda_1^2}-\frac{1}{4\Lambda_2^2}).  
\ee
For a sensible value of the mixing angle, $\theta=-19.5^{\circ}$ (i.e.  $\cos\theta=2\sqrt{2}/3$  and $\sin\theta=-1/3$), we have  $\eta=(u\bar{u}+d\bar{d}-s\bar{s})/\sqrt{3}$ and $\etp=(u\bar{u}+d\bar{d}+2s\bar{s})/\sqrt{6}$, such that Eqs. (\ref{eq:eta}) and (\ref{eq:etp}) simply reduce to 
\bea
\left.\frac{A(\bkzet)}{A(B^0\to K^0\pi^0)}\right|_{\theta=-19.5^{\circ}}  
&=&-\sqrt{\frac{2}{3}}\left[1-\zeta\right] \label{eq:lipkineta}\\
\left.\frac{A(\bkzetp)}{A(B^0\to K^0\pi^0)}\right|_{\theta=-19.5^{\circ}} 
 &=&-\sqrt{\frac{1}{3}}\left[1+2\zeta^{\prime}\right] .  
\eea
In the absence of the axial anomaly, the ideal relations given in Eq. (\ref{eq:ideal}) for
$M_{\etetp}$ imply $\zeta^{(\prime)}=1.41 (1.09)$. If in addition we assume $M_K = M_{\pi}$, then  $\zeta=\zeta^{\prime}=1$ and 
we recover the
well-known  flavour $SU(3)$ relations between the $B^0\to K^0 [0^{-+}]$ branching ratios ($\etp : \eta : \pi^0= 3:0:1$)~\cite{su3}. 
But again, the $U(1)_A$ symmetry requires physical $\etetp$ masses for a
physical $\eta-\etp$ mixing angle. Hence we have
\be
\zeta=1.29\pm 0.19, \quad \zeta^{\prime}=1.72\pm 0.26. \label{eq:nzeta}
\ee
The impact of the $U(1)$ anomaly on the $B^0 \to K^0\etp$  branching ratio
is displayed in Fig.1 for a mixing angle $\theta = -22^{\circ}$. The result
with only $SU(3)_V$ breaking corrections (a dot in Fig.1) is obtained
for $\zeta^{\prime} = 1.09$. As we have shown in Table 1, this peculiar limit
reproduces the numerical values of the hadronic matrix elements used
in the previous works. Our result which consistently includes the effect
of the $U(1)_A$ violation on the masses and mixing (a star in Fig.1) is
based on Eq. (\ref{eq:nzeta}) for $\zeta^{\prime}$. We observe a strong increase for the $B^0 \to
K^0 \etp$ branching ratio, in agreement with the data. 
 On the other hand, the relatively small $\eta$ mass induces  a small $U(1)_A$ correction to  $\zeta$, which does deviate from one mainly through the $SU(3)_V$ violation. As a result,  the $\bkzet$ branching ratio stays well below the experimental bound due to the efficient cancellation  in Eq. (\ref{eq:lipkineta}) but is, at the same time, 
extremely sensitive to
the uncertainties on $\zeta$ and $\theta$. For $\zeta = 1.29\pm 0.19$ and $\theta =
-(22 \pm 1)^{\circ}$, its value runs between $0.01 \times10^{-6}$ and $1.10\times 10^{-6}$. At this
level, contributions from the other penguin operators might be non-negligible.

In conclusion, we reconsidered  the puzzle of large $\bketp$ branching ratio  and  pointed out a missing $U(1)_A$ breaking correction to the penguin hadronic matrix element.  We estimated this correction in the framework of a low-energy effective theory of QCD in the large-$N_c$ limit. We provided the expression for all the density matrix elements relevant to the hadronic $B$ decays, in terms of physical masses and decay constants. 
We found a rather large  increase (60\%) of  
$| \la 0|\bar{s}\gamma_5s|\etp \ra|$ and a moderate decrease (10\%) of $| \la 0|\bar{s}\gamma_5s|\eta \ra|$ compared to the previous works. 
The sizable corrections for the $\etp$ 
 should not come as a surprise since its hadronic matrix element 
 vanishes in the absence of the axial anomaly. This correction
may explain why the $\bkzetp$ branching ratio is about six times larger than the $B^0\to K^0\pi^0$ one. 

%We have illustrated then that this correction plays an important role to enhance the $\bkzetp$ branching ratio since it enters as the dominant contribution. We are not expecting such an enhancement in $B\to K^*\etetp$. 
 
 \bigskip

\begin{acknowledgments}
This work was supported by the Belgian
Federal Office for Scientific, Technical and Cultural Affairs through the
Interuniversity Attraction Pole P5/27. 
\end{acknowledgments}
 
%%%%%%%%%%%%%%%%%%%%%%%%%%%%%%%%%%%%%%%%%%%%%%%%%%%

\end{document}